\documentclass{aa}  

\usepackage{graphicx}
\usepackage{txfonts}

\usepackage[colorlinks=true, linkcolor=blue, citecolor=blue, urlcolor=blue]{hyperref}

\begin{document}

\title{Dispersion relation for the linear theory of relativistic Rayleigh–Taylor instability in magnetized medium revisited}

   \author{Qiqi Jiang\inst{1,2,3}\thanks{E-mail: qiqi.jiang@desy.de}
          \and
          Guang-Xing Li\inst{1}\thanks{E-mail: ligx.ngc7293@gmail.com}
          \and
          Chandra B. Singh\inst{1}\thanks{E-mail: chandrasingh@ynu.edu.cn}
          }

   \institute{South-Western Institute for Astronomy Research, Yunnan University, Kunming, 650500 Yunnan, P.R. China
         \and
             Deutsches Elektronen-Synchrotron DESY, Platanenallee 6, 15738 Zeuthen, Germany
         \and
             Institute of Physics and Astronomy, University of Potsdam, 14476 Potsdam, Germany
             }

   \date{}
 
  \abstract
   {The Rayleigh–Taylor instability (RTI) arises at the interface between two fluids of different densities, notably when a heavier fluid lies above a lighter one in an effective gravitational field. In astrophysical systems with high velocities, relativistic corrections are necessary.
}
   {We investigate the linear theory of relativistic Rayleigh–Taylor instability (R-RTI) in a magnetized medium, where fluids can move parallel to the interface at relativistic velocities. 

}
   {We chose an "intermediate frame" where fluids on each side of the interface move in opposite directions with identical Lorentz factors $\gamma_*$. This symmetry facilitates analytical derivations and the study of relativistic effects on the instability's dynamics.

}
   {We derive the correct version of the R-RTI. We find that the instability is activated when the Atwood number $\mathcal{A}$ = $(\rho_1 h_1 - \rho_2 h_2) / (\rho_1 h_1 + \rho_2 h_2) >0$, where $rho_1$ and $\rho_2$ are densities measured in the rest from the fluids, does not contain relativistic corrections.  The relativistic effect is mostly contained in the Lorentz transformation of the gravitational acceleration $g' = g \gamma_*^{-2}$, which, combined with time dilation, leads to a much slower growth of instability ($\omega'=\gamma_*^{-1} \omega_0$), and a slightly elongated length of the unstable patch, due to weaker $g$ in the intermediate frame. Taking time dilation into account, when viewed in the rest frame of the medium, we expect the instability to grow at a much reduced rate. The analytical results should guide further explorations of instability in systems such as microquasars ($\mu$QSOs), Active galactic nuclei (AGNs),  gamma-ray bursts (GRBs), and radio pulsars (PSRs), where the apparent stability of the jet can be attributed to either the intrinsic stability (e.g. the Atwood number) or the much prolonged duration through which R-RTI can grow. 
   }
   {}

   \keywords{Instabilities --- Magnetohydrodynamics (MHD) --- Relativistic process --- Jets
               }

   \authorrunning{Jiang, Li $\&$ Singh et al.}
   \titlerunning{Relativistic Rayleigh–Taylor Instability in Magnetized Medium}
   \maketitle
%

\section{Introduction}

   Special relativity  \citep{taylor1992}, a cornerstone of 20th-century physics, plays a critical role in understanding the behavior of fluids moving at speeds close to the speed of light. This theory unifies space and time into a single entity, a concept necessary for describing relativistic astrophysical phenomena. Understanding relativistic fluids  \citep{landau1987} has become a crucial challenge with the advancement of high-energy astrophysics. Our specific topic is to study the evolution of Rayleigh-Taylor instabilities in the relativistic context.  

The Rayleigh-Taylor instability (RTI) is one fundamental macroscopic instability in nature  \citep{chandra1961, sharp}. It occurs at the interface between two different densities of fluids and becomes activated when the density gradient is in the opposite direction of the overall acceleration. The growth of the RTI can disrupt the stability of the interface, leading to subsequent mixing between two fluids, often in the form of some finger-bubble structures  \citep{ZHOU20171, Zhou2024}. The core of the theory of the RTI is the key part of other instabilities, including the Centrifugal instability (CFI)   \citep{gourgouliatos} and the Richtmyer–Meshkov instability (RMI)  \citep{ZHOU2021132838, kane, wilder}, where the effective acceleration is provided either by rotation (CFI) or by acceleration from a impulse (RMI). In an astrophysical jet, the growth of these instabilities can lead to mixing between the jet and the ambient medium, leading to the deceleration of the jet. In supernova explosions, the region between the forward shock in the interstellar medium and the reverse shock in the ejecta is RT unstable   \citep{wang}, and the growth of the instabilities can lead to mixing and energy dissipation. Numerical simulation of 3D RTI in supernova remnants studies the effect of efficient particle acceleration at the forward shock on the growth of RTI \citep{fraschetti}. The particle-in-cell (PIC) simulation of particle acceleration by magnetic RTI is performed and explains flares from the inner accretion flow onto the supermassive black hole, Sgr A*  \citep{zhdankin2023}. 

Since most astrophysical plasma is magnetized, the inclusion of the effect of the magnetic field is required. Some previous works about the magnetic RTI have been done \citep{hillier2016,hillier2018}, which mainly discuss the most unstable modes of the system in the linear stage of its evolution. \citet{2009ApJ...705.1594M} presented an approximate qualitative description related to the relation for the growth rate of rotation-induced R-RTI, accounting for the effective inertia of the poloidally magnetized inner and outer jet components. However, the results are not fully consistent (see section on length contraction paradox). \citet{Matsumoto2019} speculated about a dispersion relation for R-RTI based on physical arguments and analogy with their work on the hydrodynamic case \citep{matsumoto}. \citet{matsumoto} investigated the linear evolution theory of non-magnetized RTI at a discontinuous surface between a relativistic jet and a cocoon. But we think the results are not completely self-consistent, which is evidenced by its breakdown under a simple Lorentz transformation, because they ignore a critical assumption in their derivations, namely that the Lorentz factor measured in both frames needs to be identical, as otherwise an inconsistency of the proper length along the third dimension might render the results inconsistent.

In this paper, we first discuss some basic questions of assumptions in the analytic derivation of dimensionality reduction. Then we provide the corrected form of the dispersion relation for R-RTI in the magnetized medium by working in a frame called the "intermediate frame" where Lorentz factors of the two fluid layers are identical. 

\begin{figure}
  \centering
  \includegraphics[width=0.65\linewidth]{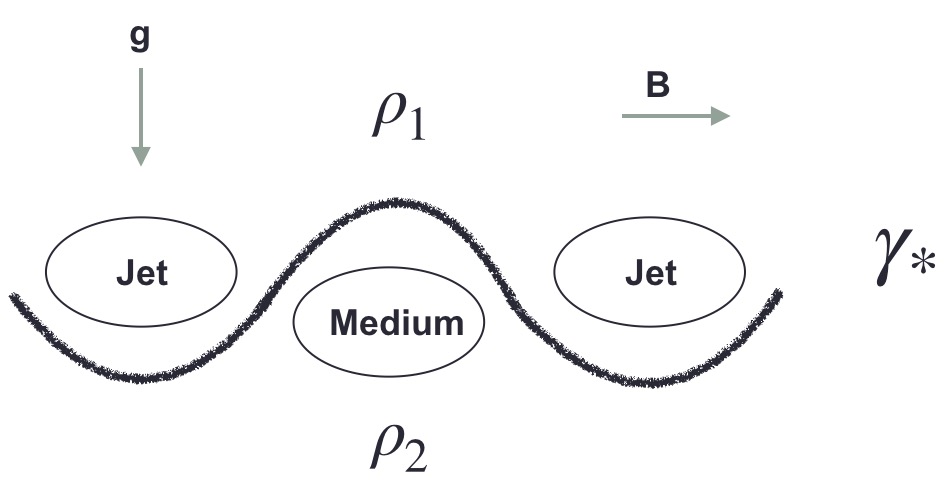} \\
  \caption{Jet cross-section in the $x$-$y$ plane\label{fig:Fig.1}.}
 \end{figure}

\begin{table}
   \centering
   \caption{List of variables \label{table:1}}
\begin{tabular}{l | l}
\hline
$\mathbf{B}$ : Magnetic field & $\mu$ : Magnetic permeability\\

$\mathbf{g}$ : Acceleration vector in $\Sigma$ & $\mathcal{A}$ : Atwood number\\

$\rho$: Rest-mass density & $k$ : Wavenumber\\

$\gamma_*$: Lorentz factor & $\omega$ : Frequency\\

$h$ : Specific enthalpy & $\mathbf{g'}$ : Acceleration vector in $\Sigma'$\\
\hline
\end{tabular}
\end{table}

\section{Length contraction paradox} \label{sec:iintermediate}
	
In most theoretical studies,  the governing magnetic
hydrodynamical equations are solved with a reduced number of dimensions. For example, in the case of the traditional RTI, one only needs to derive the dispersion relation in 1D, because the default situation is that all directions that are perpendicular to the interaction are identical. 

Considering a small size perturbation, $(l_x, l_y, l_z)$. When $l_z \gg l_x$ and $l_z \ll l_y$, the system is invariant under the translation $z' = z + \delta_z$. Due to the symmetry of 2D, we usually neglect the $z$ direction and analyze the system in the $x-y$ plane. This approach has been widely adopted in the relativistic case but may cause problems with the system in the relativistic case.  

\subsection{The theory of length contraction paradox}
\label{sec:paradox}

The \emph{length contraction paradox} refers to the case where one arrives at wrong results by solving the relativistic fluid equations with fewer dimensions. For example, solving the fluid equations in the $x$-$y$ plane when the velocity along the $z$ direction can lead to inconsistencies when the system contains relativistic velocities along the $z$ direction. 

To illustrate this inconsistency, we reconsider a perturbation of the size $(l_x, l_y, l_z)$ measured at the observer's frame at the interface between two fluids. For the analysis to be consistent, its measured length should be the same for fluids in all three directions. In the non-relativistic case, the measured size is $(l_x, l_y, l_z)$ for fluids in all possible frames, and there is no inconsistency. 

In the relativistic case, the \emph{length contraction} effect can introduce some additional constraints, which needs to be accounted for when formulating the problem.  
: Assuming that fluid 1 has velocity $(0, 0, \beta_1)$, and fluid 2 has velocity $(0, 0, \beta_2)$. This perturbation measures to $(l_x, l_y, l_z \, \gamma(\beta_1)^{-1})$ in fluid 1's frame, and  $(l_x, l_y, l_z \,\gamma(\beta_2)^{-1})$ in fluid 2's frame. The ratio between the measured length in two different frames along the $z$ direction is 
\begin{eqnarray}
    f = \frac{\gamma ({\beta_2})}{\gamma({\beta_1})}\;,
\end{eqnarray}
which is not Lorentz invariant. The relativistic motion makes it no longer possible to drop the $z$ direction from the analysis. This situation is illustrated in Fig. \ref{fig:2}. 

\subsection{ Simplifying the problem through the intermediate frame }
A direct solution for the length contraction paradox is to solve the full relativistic magnetized R-RTI equations. However, this approach is time-consuming for simulations, and tedious to perform analytically. An easy shortcut to eliminate this inconsistency is to construct a proper reference frame. In the case of relativistic RTI, suppose fluid 1 has a velocity $\beta_1$ and Lorentz factor $\gamma_1$, and fluid 2 has a velocity of $\beta_2$ and Lorentz factor $\gamma_2$, we construct a \emph{intermediate frame} where

\begin{eqnarray}
    \beta_1' &=& - \beta_{*}\\  \nonumber
    \beta_2' &=& \beta_{*}\\ \nonumber
    \gamma_1' &=& \gamma_2' = \gamma_{*} \;,
\end{eqnarray}

where $\beta_1'$  and  $\beta_2'$ are the velocities of fluid 1 and fluid 2 measured in the intermediate frame, and $\gamma$ are the corresponding Lorentz factors.

\begin{figure}
    \includegraphics[width=1.0\linewidth]{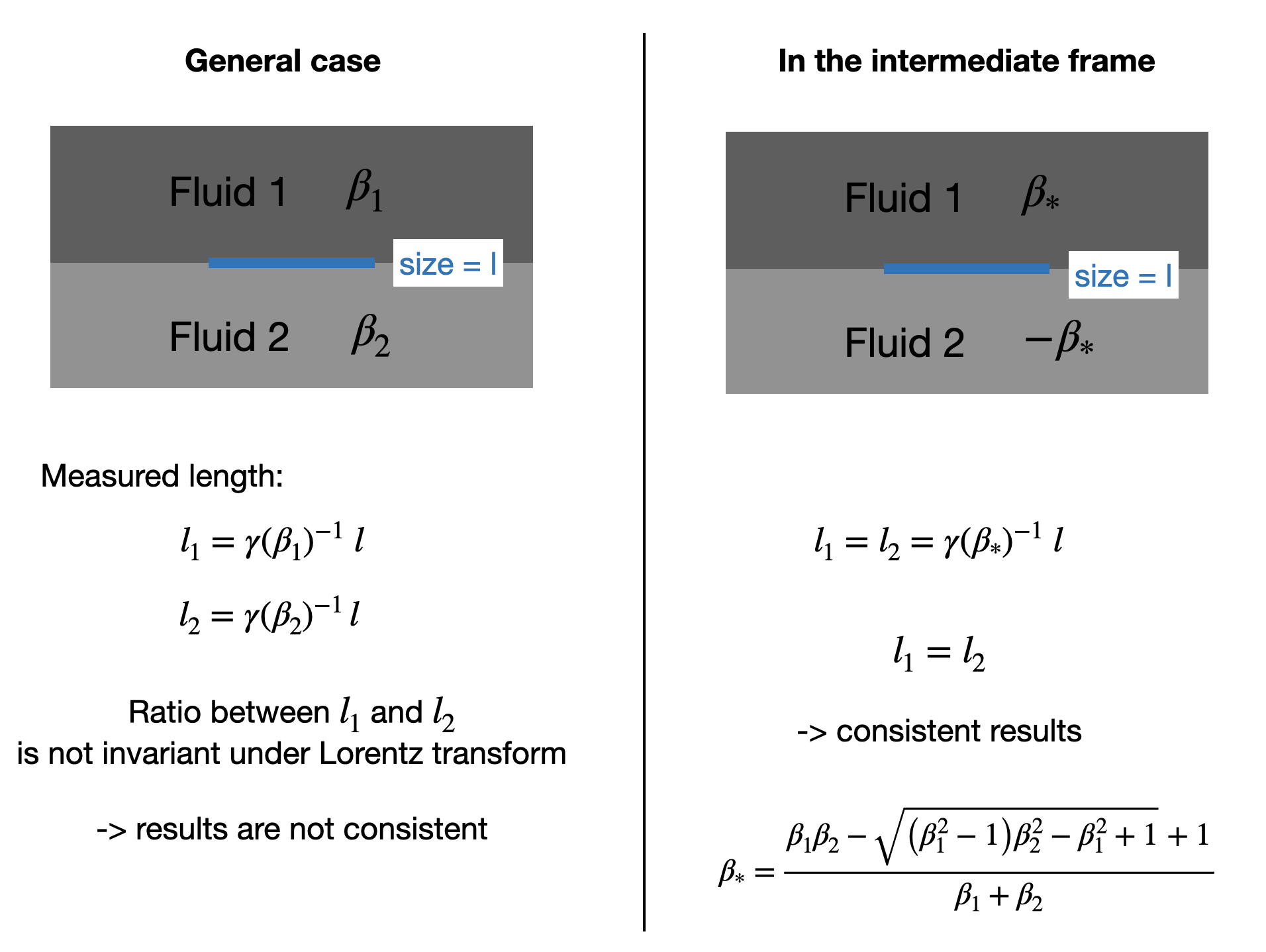}
    \caption{A illustration of how to choose the appropriate reference frame to simplify the problem in relativistic hydrodynamics. \label{fig:2}}
\end{figure}

Perturbations of size $(l_x,l_y, l_z)$ measured in the intermediate frame have the same length of $(l_x,l_y, l_z \gamma_{*}^{-1})$ seem from both the rest frame of fluid 1 and fluid 2. This smart choice of reference eliminates the inconsistency that has been discussed in \ref{sec:paradox}. An illustration of this idea can be found in Fig. \ref{fig:2}.

To derive the Lorentz factor of the intermediate frame, assuming that fluid 1 and fluid 2 have velocities of $\beta_1$ and $\beta_2$ respectively, using the formula for velocity addition, the velocity of fluid 1 measured in the intermediate frame is 
\begin{equation}
    \beta_1' = \frac{\beta_1 - \beta_{*} }{1 - \beta_{*} \beta_1} \;,
\end{equation}
and the velocity of fluid 2 measured in the intermediate frame is 

\begin{equation}
    \beta_2' = \frac{\beta_{\rm 2} - \beta_{*}}{1 - \beta_{*} \beta_2} \;.
\end{equation}
Letting $\beta_1' =  - \beta_2'$, we have
\begin{equation} \label{eq:beta:star}
 \beta_{*} = \frac{\beta_{1} \beta_{2} - \sqrt{\beta_{1}^{2}\beta_{2}^{2}  - \beta_{1}^{2} - \beta_{2}^{2} + 1} + 1}{\beta_{1} + \beta_{2}} \;,
\end{equation}
and
\begin{equation}\label{eq:gamma*}
    \gamma_{*} = \frac{1}{(1-\beta_{*}^2)^{1/2}} \;,
\end{equation}
where $\beta_{*}$ is the uniform velocity of fluids in the intermediate frame and the $\gamma_{*}$ is the corresponding uniform Lorentz factor.

A common use case of Eq. \ref{eq:beta:star} is when we stay in the same frame as one of the fluids. In this case, assuming $\beta_2 =0$, we get the Lorentz factor in the intermediate frame:
\begin{equation}\label{eq14}
    \gamma_{*} = \sqrt{\frac{\gamma_1 +1}{2}} \;.
\end{equation}
where we focus on the highly relativistic case when $\rm \gamma_1 \gg 1$. We start with this assumption of the intermediate frame to derive the dispersion relation of R-RTI in the magnetized case.

\section{Correction of dispersion relation of the R-RTI} \label{sec:result}
In Fig. \ref{fig:Fig.1}, we isolate the linear analysis of R-RTI in the magnetized medium by starting with the two-fluids box. There are two fluids of densities $\rho_1, \rho_2$ and Lorentz factor $\gamma_{*}$, separated by an interface. $\bf{g}$ is effective acceleration provided by restoring force caused by e.g. jet oscillation. The medium is magnetized with a uniform horizontal magnetic field along the $x$ axis. The meanings of the variables are summarized in Table. \ref{table:1}. 

The key observation we make is that the old dispersion relation is valid, as long as it works in the local-rest intermediate frame(See Appendix \ref{sec:appendix:derivation_revised} for the detailed derivations):
\[
\omega_{\rm local}^{2}
\;=\;
-\,g\,k \;\Bigl\{
\frac{\rho_{1}\,h_{1} \;-\; \rho_{2}\,h_{2}}
     {\rho_{1}\,h_{1} \;+\; \rho_{2}\,h_{2}}
\;-\;
\frac{\mu\,B^{2}\,k}
     {2\,\pi\,(\rho_{1}\,h_{1} + \rho_{2}\,h_{2})\,\gamma_{*}^{2}\,g}
\Bigr\}
\]
However, one may transform the dispersion of the frame where the ambient medium is at rest by applying an additional Lorentz transformation. Through the standardized procedure, the densities are already defined in the rest frames of the two fluids, and to stay in the intermediate frame correctly, we need to transform the effective acceleration $\bf{g}$, using 
\begin{equation}
    g' = \gamma_*^{-2} g\;, B' = B\;,
\end{equation}
where the value of the magnetic field $B$ is unchanged. 
Through the standard procedure of tracing the development of linear perturbation through a set of equations, including equations of mass and momentum, and using proper substitutions, we finally arrive at the following dispersion relation of relativistic magnetized RTI in the frame $\Sigma'$ moving at high speed relative to frame $\Sigma$.

\subsection{Full dispersion relation in the intermediate frame}

In the intermediate frame, the correction dispersion relation for the R-RTI is 
\begin{align}\label{eq:correct}
\omega^{2} &= -\frac{1}{\gamma_{*}^{2}}g k \left\{
\frac{\rho_{1} h_{1} - \rho_{2} h_{2}}{\rho_{1} h_{1} + \rho_{2} h_{2}} 
\right. \nonumber \\
& \quad \left. - \frac{\mu B^{2} k}{2 \pi (\rho_{1} h_{1} + \rho_{2} h_{2}) g} \right\} \nonumber \\
&= -\frac{1}{\gamma_{*}^{2}}g k \left\{ \mathcal{A} 
- \frac{\mu B^{2} k }{2 \pi (\rho_{1} h_{1} + \rho_{2} h_{2}) g}\right\} \;.
\end{align}
where $k$ is the wavenumber of the perturbation in the intermediate frame.

\subsection{The relativistic Atwood number}

We obtain the corrected relativistic Atwood number from Eq. \ref{eq:correct}
\begin{equation}\label{eq12}
    \mathcal{A} =  \frac{ \rho_{1} h_{1}-   \rho_{2} h_{2}}{ \rho_{1} h_{1} + \rho_{2} h_{2}} \;.
\end{equation}
which shows the onset of the relativistic RTI is only determined by the density contrast $\rho_{1}h_{1} - \rho_{2} h_{2}$. Here, we note that the strength of the magnetic field is small and its contribution to the specific enthalpy $h$ can be ignored. This result is consistent, especially when compared to    \citet{matsumoto}'s results where $\gamma_1^2 \rho_{1} h_{1}  -   \gamma_2^2\rho_{2} h_{2}$.

\subsection{Full dispersion relation in the jet and ambient frame}

We transform the relation to study the instabilities of the relativistic jet.  Assuming that jet have a Lorentz factor of $\gamma_{\rm jet}$, and the acceleration is caused by the expansion of the jet and applied in the rest frame of the jet, to use our result we first transform the physical quantities into the intermediate frame and back to the jet frame. By making the following substitutions
\begin{eqnarray}
    \omega_{z,  \rm jet} &=& \gamma_{*}^{-1} \omega  \nonumber \\ 
    k_{z, \rm jet} &=& \gamma_{*} k_{z}\;,
\end{eqnarray}
we arrive at the  correction dispersion relation of magnetized R-RTI in lab frame:

\begin{eqnarray}\label{eq15}
	\omega_{\rm jet} ^{2}&=&- \frac{g k_{z, \rm jet}}{\gamma_{*}^5}\left\{\mathcal{A}-\frac{\mu B^{2} k_{z, \rm jet}}{2 \pi\left(\rho_{1} h_{1}+ \rho_{2} h_{2}\right) g  \gamma_{*} }\right\}  \;.\nonumber  \\ 
	\end{eqnarray} 
where $\gamma_*$ is calculated using $\gamma_{\rm jet}$, which also holds for the rest frame of the ambient medium since they have the same $\gamma_*$

\section{Properties of the dispersion relation and relativistic effects} \label{sec:Properties of the dispersion relation and Relativistic effects}

\subsection{Condition for instability}

According to our analysis, the system becomes unstable when $\mathcal{A}>0$. According to Eq. \ref{eq12}, the instability sets in $\rho_1 h_{1} > \rho_2 h_{2}$. Whether the instability can grow or not is determined by the density contrast alone, which is identical to the non-relativistic case as discussed in  \citet{chandra1961}. 
\textit{Motion parallel to the interface does not affect the critical condition for the onset of the instability.} 

\subsection{Critical wavelength}
Neglecting the magnetic field, we get the unstable modes from Eq. \ref{eq:correct},
\begin{equation}
  k <  k_{\rm crit, non-relativistic} =   \frac{2 \pi g (\rho_{1} h_{1}-   \rho_{2} h_{2})} {{\mu} B^2}  \;,
\end{equation}
which contains no explicit relativistic correlations.

\subsection{Growth rate}
In the intermediate frame, neglecting the magnetic field,  the growth rate $\sigma$ of the
instability is given as 
\begin{equation}
    \sigma = \sqrt{g' k \mathcal{A}} = \gamma_*^{-1} \sqrt{g k \mathcal{A}}   \;,
\end{equation}

where the relativistic effect is contained in the Lorentz factor $\gamma_{*}$, leading to a slower growth rate.

When observed in the frame of the medium, the growth rate is
\begin{equation}
    \sigma_{\rm medium} = \gamma_{\rm jet}^{-1} \sigma =\gamma_{*}^{-2} \sqrt{g k \mathcal{A}}   \;,
\end{equation}
which can be even slower.

\section{Astrophyscial implications}
\begin{figure}
  \centering
  \includegraphics[width=0.86\linewidth]{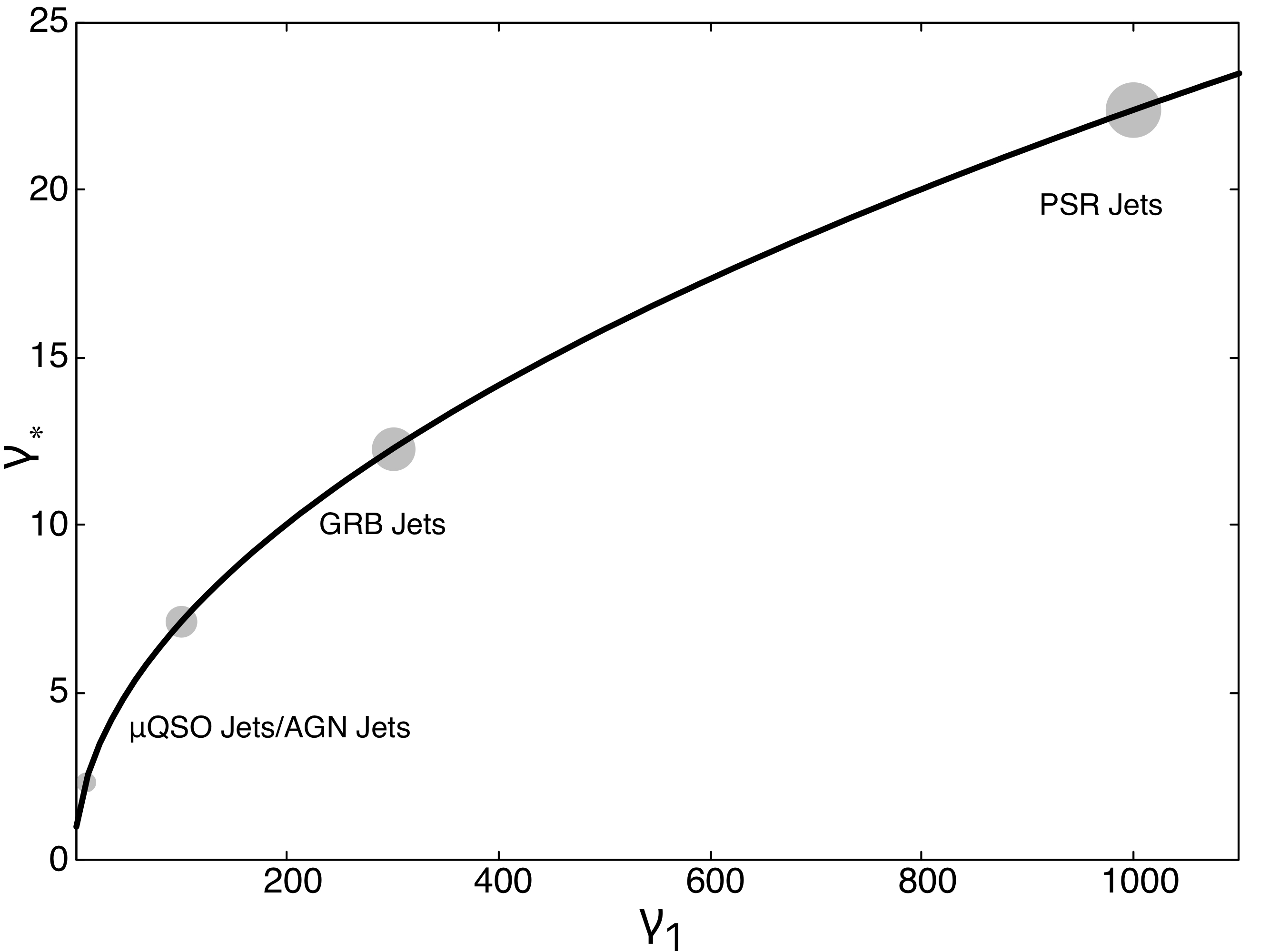} \\
  \caption{Theoretical Lorentz factor values for jets from different astrophysical sources in the intermediate frame.
  \label{fig:Fig.3}}
 \end{figure}

\begin{figure}
  \centering
  \includegraphics[width=1.0\linewidth]{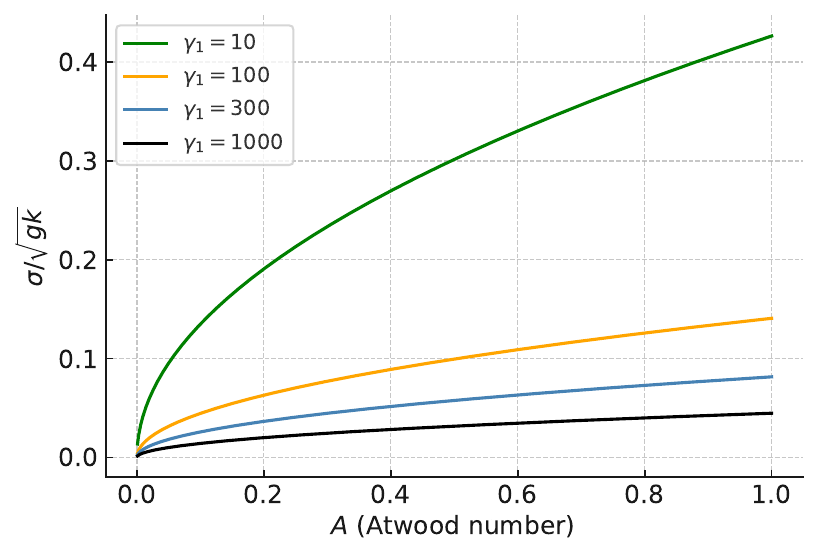} \\
  \caption {The dimensionless growth rates of the R-RTI. The green, yellow, blue, and black lines show the analytical relation between the Atwood number and dimensionless growth rates $\sigma / {\sqrt{gk}}$ with different Lorentz factors $\gamma_1=10,100,300,1000$, respectively.
 \label{fig:Fig.5}}
 \end{figure} 

 Our new analytical results of R-RTI above have indeed been clearly and correctly achieved in the understanding of the properties of the evolution of RTI in relativistic regimes. We think it's very interesting to apply it to the jets observed in different classes of astrophysical sources, such as active galactic nuclei(AGN), microquasars($\mu$QSO), gamma-ray bursts(GRB), and so on. This made it possible to explain the observation results using the theoretical analysis. Thus, we plot the theoretical Lorentz factors of different jets in the intermediate frame according to equation \ref{eq14} in Fig. \ref{fig:Fig.3}. For AGN jets, the mean Lorentz factor ranges from about 10-100  \citep{hovatta}. For GRB jets, $\gamma \sim 10^2$  \citep{rees}, here we take about 300  \citep{beskin}. For jets in microquasars and radio pulsars, the mean Lorentz factor is about 1000  \citep{beskin}. Fig. \ref{fig:Fig.3} shows that actually, the relativistic jets with different high Lorentz factors have relatively small relativistic effects in the intermediate frame. Both AGN jets and GRB jets have Lorentz factor sizes below 15 in the intermediate frame, which means it's difficult for jets to reach the high Lorentz factor in the intermediate frame. Besides, the horizontal magnetic field here plays a role in suppressing the growth of instability.

We compare the growth rates of R-RTI in the jets from different sources. Fig.\ref{fig:Fig.5} shows the high growth rates for jets with large Lorentz factors, which means the instability grows fast in the superfast jets. The density contrast between the jet and ambient medium is also an important reason for the growth rates, thus, the effect of the density ratio of jet and medium $\eta$ on the growth of instability is extensively studied in many numerical simulations.

\section{Conclusions and discussions}
The relativistic Rayleigh–Taylor Instability (R-RTI) is expected to occur in astrophysical jets from different sources,  e.g. AGNs and GRBs. We present the correct form of the linear theory of the evolution of relativistic R-RTI in the magnetized medium.
By introducing the concept of an "intermediate frame" where the two fluids move toward opposite directions with common Lorentz factors, we propose a procedure to address the inconsistencies in the our derivations. This intermediate frame is a convenient frame of choice to understand the relativistic effects. 

Whether R-RTI can develop is determined by the relativistic Atwood number. We find that under relativity, the Atwood number retains the original form
\begin{equation}\nonumber
    \mathcal{A} =  \frac{ \rho_{1} h_{1}-   \rho_{2} h_{2}}{ \rho_{1} h_{1} + \rho_{2} h_{2}} \;.
\end{equation}
contains no apparent relativistic correlations. Thus, the development of R-RTI is determined only by density contrast. 

The relativistic effects leads to lower growth rates of the instability in the intermediate frame, and a much lower rate in the rest frame of the medium, due to the fact that gravitational acceleration is reduced by the Lorentz factor squared

The astrophysical jets are complex systems where the magnetic field can organize in several different ways, and other instabilities, such as the Kelvin–Helmholtz instability(KHI) can also occur. Nevertheless, our results of the development of R-RTI in the linear regime set up the function upon which understanding of these complex systems can be established. The must reduced growth rate also may explain the apparent stability of many of the astrophysical jets.

Future work on relativistic fluid dynamics will be crucial for understanding more complex astrophysical systems. As observational capabilities improve, incorporating non-linear effects and exploring interactions between different instabilities will be key. This ongoing research will deepen our understanding of phenomena such as black hole accretion, cosmic jets, and compact object formation, cementing the importance of relativistic hydrodynamics in astrophysics.

\begin{acknowledgements}
      QJ and CBS are supported by the National Natural Science Foundation of China (NSFC) under grant No. 12073021.
GXL is supported by the NSFC under grants No. 12273032 and 12033005. We thank Prof. Martin Pohl for proofreading
this paper and wise comments.
\end{acknowledgements}

%
%

\appendix
\section{Revised Derivation of the Magnetized Relativistic Rayleigh-Taylor Instability}\label{sec:appendix:derivation_revised}

In this appendix, we present a revised derivation of the dispersion relation for the magnetized relativistic Rayleigh-Taylor instability (R-RTI). We consider two fluids separated by an interface at $y=0$, with rest-mass densities $\rho_1$ and $\rho_2$, and specific enthalpies $h_1$ and $h_2$ in Fig. \ref{fig:Fig.1}. A uniform horizontal magnetic field $\mathbf{B}$ is imposed along the $x$-axis, and an effective gravitational acceleration $\mathbf{g}$ is assumed to act along $-y$. We note that in a relativistic setting, the vector $\mathbf{g}$ transforms under Lorentz boosts, so the form $\mathbf{g}=(0,-g,0)$ holds strictly in one particular frame only. Below, we first derive the dispersion relation and then discuss the implications of a Lorentz transformation along $x$.

\subsection{Basic equations and linearization in the local rest frame}

Following the standard approach to relativistic magnetohydrodynamics (MHD), we write the continuity, momentum, and induction equations in an ideal MHD limit. In the chosen frame $\Sigma$, both fluids share the same Lorentz factor $\gamma_{*}$, and the gravitational acceleration is $\mathbf{g} = (0,\, -g,\, 0)$. The relevant equations are
\begin{equation}\label{eqa1}
\frac{\partial  (\gamma_{*} \rho)}{\partial t}
\;+\;\nabla \cdot(\gamma_{*} \rho \,\mathbf{v}) \;=\;0\,,
\end{equation}
\begin{equation}\label{eqa2}
\gamma_{*}^{2} \,\rho\,h 
\Bigl[\frac{\partial \mathbf{v}}{\partial t}
\;+\;(\mathbf{v} \cdot \nabla)\,\mathbf{v}\Bigr]
\;=\;-\nabla P
\;-\;\frac{\mathbf{v}}{c^{2}}\;\frac{\partial P}{\partial t}
\;+\;\gamma_{*}^{2} \,\rho\,h \,\mathbf{g}
\;+\;\mu\,(\mathbf{J} \times \mathbf{B})\,,
\end{equation}
\begin{equation}\label{eqa3}
\nabla \cdot \mathbf{B}\;=\;0\,,
\end{equation}
\begin{equation}\label{eqa4}
\frac{\partial \mathbf{B}}{\partial t}
\;=\;\nabla \times(\mathbf{v} \times \mathbf{B})\,,
\end{equation}
where $\rho$ is the rest-mass density, $P$ is the total pressure (gas + magnetic), $P=P_{\rm gas}+\frac{B^{2}}{2 \gamma_{*}^{2}}+\frac{(v \cdot \mathbf{B})^{2}}{2}$, $h$ is the total specific enthalpy, $h=h_{\rm gas}+\frac{B^{2}}{\gamma_{*}^{2} \rho}+\frac{(v \cdot \mathbf{B})^{2}}{\rho}$($h_{\rm gas}=1+\frac{\Gamma}{\Gamma-1} \frac{P}{\rho c^{2}}$ ), $\mathbf{v}$ is the velocity, $\mu$ is the permeability, and $\mathbf{J}$ is the current density. The interface between the two fluids is at $y=0$, with $(\rho_1,\, h_1)$ for $y>0$ and $(\rho_2,\, h_2)$ for $y<0$. A uniform magnetic field $\mathbf{B}$ points along $x$.

\subsubsection{Linear perturbations}

We decompose each physical quantity into an equilibrium value plus a small perturbation, e.g.,
\[
\rho \;\to\; \rho + \delta\rho,\quad
P \;\to\; P + \delta P,\quad
\mathbf{B} \;\to\; \mathbf{B} + \delta \mathbf{B},\quad
\mathbf{v} \;\to\; \delta \mathbf{v}
\]
We then substitute these into Eqs.~(\ref{eqa1})--(\ref{eqa4}), neglecting higher-order terms. We have a set of perturbed equations:
\begin{equation}\label{eqa6}
\frac{\partial}{\partial t}(\gamma_{*} \delta \rho)+\gamma_{*} \rho \nabla \cdot \delta \mathbf{v}=0\;,
\end{equation}

\begin{flalign}\label{eqa7}
\gamma_{*}^{2} \rho h [\frac{\partial}{\partial t}\delta\mathbf{v} + (\mathbf{v} \cdot \nabla)\delta\mathbf{v}]= \nonumber\\ -\nabla \delta P + \gamma_{*}^{2}\rho h \mathbf{g} + \frac{4 \pi}{\mu}[(\nabla\times\mathbf{B})\times\delta\mathbf{B}+(\nabla\times\delta\mathbf{B})\times\mathbf{B}]\;,
\end{flalign}

\begin{equation}\label{eqa8}
\nabla \cdot \delta \mathbf{B}=0 \;,
\end{equation}

\begin{equation}\label{eqa9}
\frac{\partial}{\partial t}(\delta \mathbf{B})=\nabla \times(\delta \mathbf{v} \times \mathbf{B}) \;.
\end{equation}
We expand it in the x-y plane and get a set of linearized equations:

\begin{equation}\label{9}
\frac{\partial}{\partial t}(\gamma_{*} \delta \rho)+\gamma_{*} \rho (\frac{\partial}{\partial x} \delta v_x+\frac{\partial}{\partial y} \delta v_y)=0\;,
\end{equation}

\begin{flalign}
\gamma_{*}^{2} \rho h [\frac{\partial}{\partial t}\delta {v_x} + {v_x} \frac{\partial}{\partial x}\delta {v_x} + {v_y} \frac{\partial}{\partial y} \delta {v_x}] =\nonumber \\ 
 -\frac{\partial}{\partial x }\delta P +\frac{4\pi}{\mu}[-\frac{\partial {B_x}}{\partial y}\delta {B_y} + \frac{\partial \delta{B_y}}{\partial x}\cdot {B_y}]\;,
\end{flalign}

\begin{flalign}
\gamma_{*}^{2} \rho h [\frac{\partial}{\partial t}\delta {v_y} + {v_x} \frac{\partial}{\partial x}\delta {v_y} + {v_y} \frac{\partial}{\partial y} \delta {v_y}]=  \nonumber \\ -\frac{\partial}{\partial y }\delta P + \gamma_{*}^{2}\rho h \mathbf{g}+\frac{4\pi}{\mu}[\frac{\partial {B_y}}{\partial x}\delta {B_x} - \frac{\partial {B_x}}{\partial x}\cdot \delta{B_y}]\;,
\end{flalign}

\begin{equation}
\frac{\partial}{\partial x} \delta {B_x}+\frac{\partial}{\partial y} \delta {B_y} =0 \;,
\end{equation}

\begin{equation}\
\frac{\partial}{\partial t}(\delta {B_x})=\frac{\partial}{\partial y}(v_y B_x - v_x B_y) \;,
\end{equation}

\begin{equation}\label{14}
\frac{\partial}{\partial t}(\delta {B_y})=\frac{\partial}{\partial x}(v_x B_y - v_y B_x) \;,
\end{equation}
Then we take normal-mode perturbations of the form 
\[
\delta \rho, \delta P, \delta \mathbf{B}, \delta \mathbf{v} \propto \exp i (k x- \omega t)\;,
\]
Insert these small perturbations into the linearized Eqs. \ref{9}-\ref{14}, we obtain

\begin{equation}\label{16}
-i \omega \gamma_{*} \delta \rho + \gamma_{*} \rho (ik \delta v_x + v_y D_y \delta v_y)=0 \;,
\end{equation}

\begin{flalign}
\gamma_{*}^{2} \rho h [-i \omega \delta v_x + ik v_x \delta v_x + v_y D_y \delta v_x]  = \nonumber \\ -ik \delta P + \frac{4\pi}{\mu}[-D_y B_x \delta B_y + ik \delta B_y B_y] \;,
\end{flalign}

\begin{flalign}
\gamma_{*}^{2} \rho h [-i \omega \delta v_y + ik v_x \delta v_y + v_y D_y \delta v_y]  = \nonumber \\-D_y \delta P + \gamma_{*}^{2} \rho h \mathbf{g} +\frac{4\pi}{\mu}[D_x B_y \delta B_x - D_x B_x \delta B_y] \;,             
\end{flalign}

\begin{equation}
ik \delta B_x + D_y \delta B_y =0 \;,
\end{equation}

\begin{equation}
-i \omega \delta B_x = D_y(v_y B_x - v_x B_y) \;,
\end{equation}

\begin{equation}\label{21}
-i \omega \delta B_y = D_x(v_x B_y-v_y B_x) \;.
\end{equation}

Here, we have $D_{x}=\frac{\partial}{\partial x}$ and $D_{y}=\frac{\partial}{\partial y}$. Eliminating $\delta\rho$, $\delta P$, and $\delta\mathbf{B}$ yields a single differential equation for $\delta v_y$. In each uniform region ($y>0$ or $y<0$), $\delta v_y$ satisfies
\[
\bigl(\tfrac{\partial^2}{\partial y^2} \;-\; k^2\bigr)\,\delta v_y \;=\;0,
\]
whose solutions are exponentials decaying or growing away from the interface. We impose continuity across $y=0$ and integrate across the discontinuity in density/enthalpy. This yields the local-frame dispersion relation
\begin{equation}\label{eq:DR_sigma}
\omega_{\rm local}^{2}
\;=\;
-\,g\,k \;\Bigl\{
\frac{\rho_{1}\,h_{1} \;-\; \rho_{2}\,h_{2}}
     {\rho_{1}\,h_{1} \;+\; \rho_{2}\,h_{2}}
\;-\;
\frac{\mu\,B^{2}\,k}
     {2\,\pi\,(\rho_{1}\,h_{1} + \rho_{2}\,h_{2})\,\gamma_{*}^{2}\,g}
\Bigr\},
\end{equation}
where $\omega_{local}$ is the mode frequency, $k$ is the wavenumber along $x$, $\gamma_{*}$ is the (common) Lorentz factor of the two fluids, and $B$ is the uniform magnetic field strength. Instability arises when $\omega^2<0$, implying $\operatorname{Im}(\omega)\neq 0$.

From the above, one defines the growth rate $\sigma = \operatorname{Im}(\omega)$ and a dimensionless Atwood number 
\[
\mathcal{A} 
\;=\;
\frac{\rho_{1}\,h_{1} \;-\; \rho_{2}\,h_{2}}
     {\rho_{1}\,h_{1} \;+\; \rho_{2}\,h_{2}},
\]
so that $\sigma/\sqrt{g\,k} = \sqrt{\mathcal{A}}$ in the limit $B=0$. Thus, in the non-magnetized case, the relativistic RTI growth rate reduces to a form analogous to the classical one.

\subsection{Lorentz transformation along \texorpdfstring{$x$}{x} axis: Effects on \texorpdfstring{\boldmath$\mathbf{g}$}{g}}
\label{sec:B}
The derivation above is carried out in the specific frame $\Sigma$, where both fluids share the same Lorentz factor $\gamma_{*}$, and $\mathbf{g} = (0,\, -g,\, 0)$. However, in special relativity, if one boosts the system along the $x$ direction by some velocity $v$ to a new frame $\Sigma'$, both the fluid velocities and the gravitational acceleration will change. Hence, to write down the RTI dispersion relation in $\Sigma'$, one must replace 

\[
(\rho_i,\,h_i,\,\gamma_{*},\,\mathbf{B},\,\mathbf{g})
\;\longrightarrow\;
(\rho_i,\,h_i,\,\gamma_{*},\,\mathbf{B},\,\mathbf{g}'),
\]
where 
\[
\mathbf{g}' 
\;=\;
\bigl(0,\,-\tfrac{g}{\gamma_*^{2}},\,0\bigr)\;,
\]
Since quantities like $\rho_i$, $h_i$, and $\gamma_i$ are already defined in their reset frames, the magnetic field $B$ in invarient under the Lorentz transformation and only the relativistic effects on the gravitational acceleration needs to be considered.

\end{document}